# Magnetic relaxation in pulse-magnetized high-temperature superconductors.

Kartamyshev A.A., Krasnoperov E.P., Kuroedov Yu.D
RRC Kurchatov Institute, 123182, Moscow, Russia

Nizhelskiy N.A., Poluschenko O.L.
MGTU Bauman, 105005, Moscow

Abstract.

*A magnetic filed relaxation at the center of a pulse-magnetized single-domain Y-Ba-Cu-O superconductor at 78K has been studied. In case of a weak magnetization, the magnetic flux density increases logarithmically and normalized relaxation rate defined as $S=-d(lnB)/d(lnt)$ is negative ($S=-0.037$). When an external magnetic field magnitude increases, the relaxation rate first decreases in absolute value, then changes sign (becomes positive, $S>0$) and after reaching some maximum finally reduces to a very small value. Non-monotonous dependence of $S$ vs $H_a$ is explained by a non-homogeneous local temperature distribution during a pulse magnetization.*

**Introduction**.
Textured high-temperature superconductors (HTS) of Re(Y)-Ba-Cu-O family appear to be very promising materials for manufacturing superconducting magnets with very large magnitudes of trapped magnetic fields [1, 2]. For example, trapping a field as high as 17 T between two 26.5-mm diameter Y-Ba-Cu-O disks at 29K has been demonstrated recently [3]. A comprehensive review of synthesis, characteristics and applications of high-quality HTS suitable for applications in superconducting magnets can be found in [4]. Recently significant research efforts have been directed towards using a pulse magnetization of HTS instead of an isothermal one [5]. An obvious advantage of the pulse magnetization with a typical pulse duration in order of 1 to 100 msec is a smaller amount of energy needed to create a magnetic field of a desired magnitude. There is, however, a complication that needs to be considered when using this method: an electromagnetic heating of a superconductor during the pulse magnetization plays much bigger role than during a slow magnetization because there is not enough time for the temperature to reach an equilibrium distribution within the superconducting volume. For example, a thermal time constant of a HTS sample at 78K with characteristic dimensions in order of 10 mm would be in order of a few seconds, which is much large than the pulse duration. This circumstance allows one to analyze the pulse magnetization as an adiabatic process. A negative effect of the heating during a single-pulse magnetization can be reduced by using a multi-pulse method. Using a combination of the multi-pulse magnetization and a step-by-step temperature decrease has been shown to allow trapping magnetic fields as high as 4.5T at 30K [6].

One of the key performance characteristics of any magnet is a long-term stability of its magnetic field. Here superconducting magnets have a fundamental limitation due to a phenomenon known as a thermally-activated magnetic flux creep. According to Kim-Andersen theory a thermally-activated magnetic flux creep in type II superconductors [7], supercurrents and magnetization decrease logarithmically in time as

$$J=J_{co}[1-kT/U_o \ln(t/t_o)] \quad (1)$$

where $U_0$ is a height of a potential barrier to magnetic vortex movements and $J_{c0}$ is a critical current density in the absence of thermal activation. A logarithmic derivative $S=-d\ln J/d\ln t = kT/U_o$ characterizes a magnetic relaxation or the rate of a magnetic creep. The minus sign is chosen to have positive S when the current decreases. Although there are many models describing magnetic relaxation by considering Josephson contacts, Bean-Livingston surface barriers, crystallographic anisotropy, etc., the classical thermo activation model remains the most effective one for analyzing the magnetic flux creep in Re(Y)-Ba-Cu-O monocrystals (at least at temperatures above 50K). Usually the magnetic relaxation is investigated either under FC conditions with the external field being reduced to zero after magnetization, or under ZFC with the external magnetic field sustained. In ideal Re(Y)-Ba-Cu-O crystals, the relaxation rate would be independent of the magnetic history and would depend only on $J_c$ [7]. If the external field is switched off, the magnetization would always decreases with time in this case.

So far the magnetic flux relaxation after the pulse magnetization has not been studied as extensively as the relaxation after the isothermal magnetization. One example of such a study can be found in [8], where the magnetic flux dynamics in a single-domain Y-Ba-Cu-O disk at 78K was investigated. Three Hall-effect sensors were placed on the disk face: one at the center (disk axis), one at equal distances from the center and the edge of the disk and one on the edge. These Hall sensors were used to measure the magnetic field B first when the external magnetic field was applied and then $\approx 10^3$ sec later. The results (see fig. 11 in paper [8]) indicated a smaller relaxation rate in case of a pulse magnetization with a 1T magnetic field than in case of an isothermal magnetization (FC, ZFC). Also, it was shown that S decreases when the external magnetic field increases. Another source [9] suggests that during short periods of time after a magnetization pulse relaxation curves B(t) depend on the external magnetic field parameter of pulse; however when more time passes this dependence disappear. Because of this observation, it was interesting to repeat the experiment described in [8] on a larger time scale.

In the present work, results of studies of the magnetic relaxation following a pulse field magnetization of textured (quasi monocrystal) Y-Ba-Cu-O disks at 78K are presented. The studies have been conducted for different magnetization regimes.

**Experimental setup**.

For the purpose of the experiment, quasi single-domain crystal Y-Ba-Cu-O cylinders were fabricated using the top-seeded melt-growth process with an elongated $GdBa_2Cu_3O_y$ crystal used as a seed [10]. Gd-Ba-Cu-O was preferred to Sm-Ba-Cu-O as the seed material because of its lower melting temperature. The main advantage of using an elongated seed is that it produces a long crystal on the surface of a sample at the very beginning of the crystallization process, which subsequently facilitates a persistent growth of a well-oriented crystal within a large volume. The crystal seed was placed on the '001' crystal plane. When a Y-Ba-Cu-O sample with a seed on top is held at the crystallization temperature, a crystal starts forming right under the seed and then continues growing through the entire volume of the sample. This method allows fabricating disk samples up to 48 mm in diameter. Small disks with 15mm diameter and 11mm thickness used in the experiment were cut from so-obtained pellets. Crystal axis **c** was oriented along the disk axis. A couple of disks were fixed in a non-conducting holder with a 1mm gap between them. A Hall sensor placed in the gap at the center of the disk assembly allowed measurement of the magnetic flux density. The resulting flux density measurements were much more meaningful than if a single disk were used because of a much smaller demagnetization factor [5, 9]. The measurements were conducted during the magnetization as well as up to $\approx 10^4$ sec after it. The Hall sensor sensitivity was 74 mV/T. The pulse magnetic field system used for the magnetization is described in [11]. The magnetization coil had two sections (quasi Helmholtz system) with 80 mm inner diameter. The field homogenous throughout the sample was better than $10^{-3}$. The magnetic pulse had a half sinusoidal form and duration 10 ms. Sample was contained in of a thermostat filled with liquid nitrogen.

The experimental method combining pulse and DC measurements was similar to the one described in [8]. During the magnetic pulse and the following 100 ms readings from the Hall sensor were registered using digital oscilloscope. For longer periods of time, a digital multimeter with 0.1 μV accuracy and 2 Hz sampling frequency was used.

**Results.**

At the beginning of the experiment, isothermal magnetization curves, trapped fields in the FC mode and the magnetic relaxation in the 50 – 80 K temperature range were studied with the magnetizing field being generated by a superconducting solenoid. The magnetic field trapped by the disk assembly under FC conditions at 50 K after the solenoid was turned off was found to be around 7 T. The trapped magnetic field magnitude was decreasing with temperature and at T=75K it was around 1.4T. At T=80K the trapped magnetic field was further reduced to 1T. Because no flux jumps had been observed during the external field reduction, it was concluded that superconducting currents flowed throughout the entire cross section of the sample. The magnitudes of these currents can be easily estimated using Bean's model (as in [12]), which assumes a constant critical current density in superconductor equal to $J_c$. This assumption results in the following estimate of the magnetic flux density trapped in a superconducting cylinder of radius R

$$B(r)=0.4\pi(R-r) J_c \qquad (2).$$

Based on (2), the trapped magnetic flux density increases linearly from zero at the edge of disk up to a maximum value $B_o$ at the disk center. Fig. 1 shows a few typical distributions B(r).

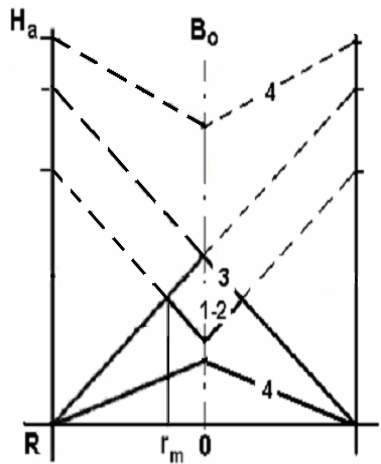

Fig. 1. Magnetic flux density in a superconducting cylinder calculated using an idealized Bean's model. Dotted lines – magnetic flux density distribution in an external field, solid lines - distribution of the trapped magnetic flux density B(r): 1-2 – "undermagnetized" state, 3 – max flux density state, 4 – maximum flux density under a higher temperature than in 2 and 3.

The dotted lines represent B measured in an external field $H_a$, whereas the solid lines represent the magnetic flux density trapped after the external field is reduced to zero. The slope $dB/dr = -0.4\pi J_c(T)$ is determined by the critical current density $J_c$, which is assumed to be a function of the temperature only (the magnetic field dependence is neglected). For the samples used in the experiment, the critical current density was estimated using equation (2) to be $J_c \approx 2 \cdot 10^4 A/см^2$ at T=75K. The applicability of Bean's model and the linearity of the magnetic field vs. radius dependence have been demonstrated in [3] under the liquid nitrogen temperature.

On the next stage of the experiment, pulsed field magnetizations under the liquid nitrogen temperature (T=78K) were studied. Prior to each magnetic pulse a sample was first heated up to more than 120K and then cooled down to T=78K (ZFC). The dependence of a magnetic flux density $B_o(H_a)$ vs. magnitude of the magnetizing pulse $H_a$ is shown in Fig. 2. From this figure one can see that the field at the center of the disk assembly remains zero as long as the external magnetic field is less than $\mu H_a \approx 1.7T$. When the external magnetic field increases beyond this limit, the total magnetic flux density first reaches a maximum and then decreases sharply. The observed $B_o(H_a)$ dependence is similar to the one presented in [13] for Sm-Ba-Cu-O.

A dependence of the magnetic flux density vs. the external field in case of the isothermal magnetization is shown in fig. 2 (dotted line). Here the external field (after ZFC) was slowly increased up to $H_a$ and then slowly brought down to zero. This linear dependence was calculated based on the earlier estimate of the critical current density and equation (2). According to Bean's model [12], the total magnetic field at the center of the disks becomes non-zero when the external magnetic field exceeds the maximum field that can be trapped in the sample, i.e. when $\mu H_a = B_o$. The maximum field will be trapped when the external field is doubled ($\mu H_a = 2B_o$). This case is marked as 3 in fig. 1. The curve 1-2 in fig. 1 illustrates the case when the sample is "under-magnetized", i.e. when the magnetic flux density at the center of the sample is lower than the maximum trapped flux density. After the external magnetic field is turned off with the sample being "under-magnetized", the maximum residual magnetic field would be equal to $J_c \cdot (R - 2r_m)$. This time the maximum would occur at some non-zero radius $r_m$ ($0 < r_m < R$).

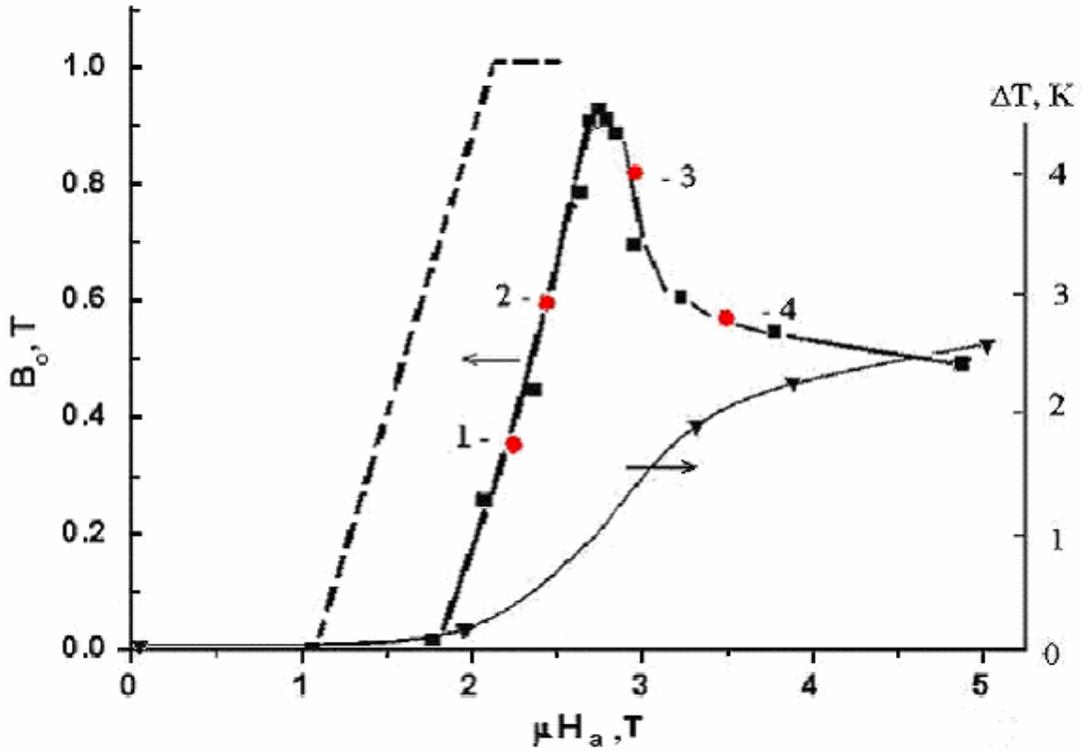

Fig. 2. Magnetic flux density (● and ■ – left axes) and temperature jump (▼– right axes) vs. magnetizing field magnitude. Dotted line is a theoretical prediction based on Bean's model of the critical state at $B_{o\,max}$ =1.1 T.

Bean's model works well also for a pulse magnetized single-domain Y-Ba-Cu-O, in which case the distribution B(r) was also found to be close to linear [14, 15]. Likewise, a linear dependence of the trapped field vs. the external field can be expected. Fig. 2 indicates that the external magnetic field in this case needs to be by approximately 0.7 T higher than in case of the isothermal (slow) magnetization. This difference was explained by a time delay between the moments when the external field was applied and when it penetrated into a superconductor. With the field penetration delay taken into account, the rising part of the $B_o(H_a)$ dependence may be well described by an idealized Bean's model [12]. Uphill part of the $B_o(H_a)$ dependence shown in Fig. 2 (when $\mu H_a > 2,7T$) appears due to heating a superconductor during pulse magnetization [13]. An increasing temperature leads to a degradation of the critical current density, which, in turn, leads to a decrease of the trapped magnetic flux density $B_0$. This model has been validated in [16], where a numeric calculation of the total magnetic flux in a bulk sample vs. pulse magnitude was carried out and its results were found to be in a good agreement with experimental data. However, because the total flux was an integral of the magnetic field $\Phi = \int B ds$, the slope of the $\Phi(H_a)$ curve was quite small. The dependence $B(H_a)$ shown in fig. 2 has a narrower peak than in [13], because we used a shorter magnetization pulse resulting in stronger localized heating.

To estimate heating from a pulse magnetization, the temperature at the center of the disk was measured with a differential thermocouple. One of the thermocouple junctions was glued to a surface of the superconductor and additionally pressed again it through a thermally insulating spacer. The other junction was submerged into the liquid nitrogen. The observed dependence of the temperature vs. pulse magnitude is shown in fig. 2 (symbols ▼, right scale of ordinate). With a 2 Hz sampling rate and the disk thermal time constant being in order of several seconds, the accuracy of the initial temperature estimate after extrapolation was close to $\Delta T/T \approx 20\%$.

Heating the sample could be analyzed as an adiabatic process because the duration of a magnetization pulse was much smaller than the disk thermal time constant. The fastest temperature relaxation was in the (ab) crystal plane, which was the plane normal to the disk axis. The time constant of the thermal relaxation there was $\tau=R^2/D_h=5$ sec, where $D_h =0.1 cm^2/sec$ is the heat diffusion coefficient [17]. This time was two orders of magnitude higher than the pulse duration (10 ms). The critical current density $J_c$ at elevated temperatures is known to be lower than at 78K:

$$J_c(78) - J(T) = |\partial J_c/\partial T| \cdot \Delta T \quad (3)$$

With the pulse magnitude $H_a$ being approximately 5 T, the temperature increase at the center of sample exceeded 2.7 K (Fig. 2). Based on the available isothermal magnetization data, $dB_{tr}/dT$ in this temperature range was $\approx -0.11$ T/K. Therefore, hearting the superconductor was expected to decrease the trapped magnetic flux density by more than 0.3 T from the maximum value. This is in agreement with the experimental data shown in fig. 2.

The long-term magnetic relaxation was studied for different magnetization conditions. Fig.3 shows normalized values B(t)/B(0) for states noted by (*) and labeled 1 through 4 in fig. 2.

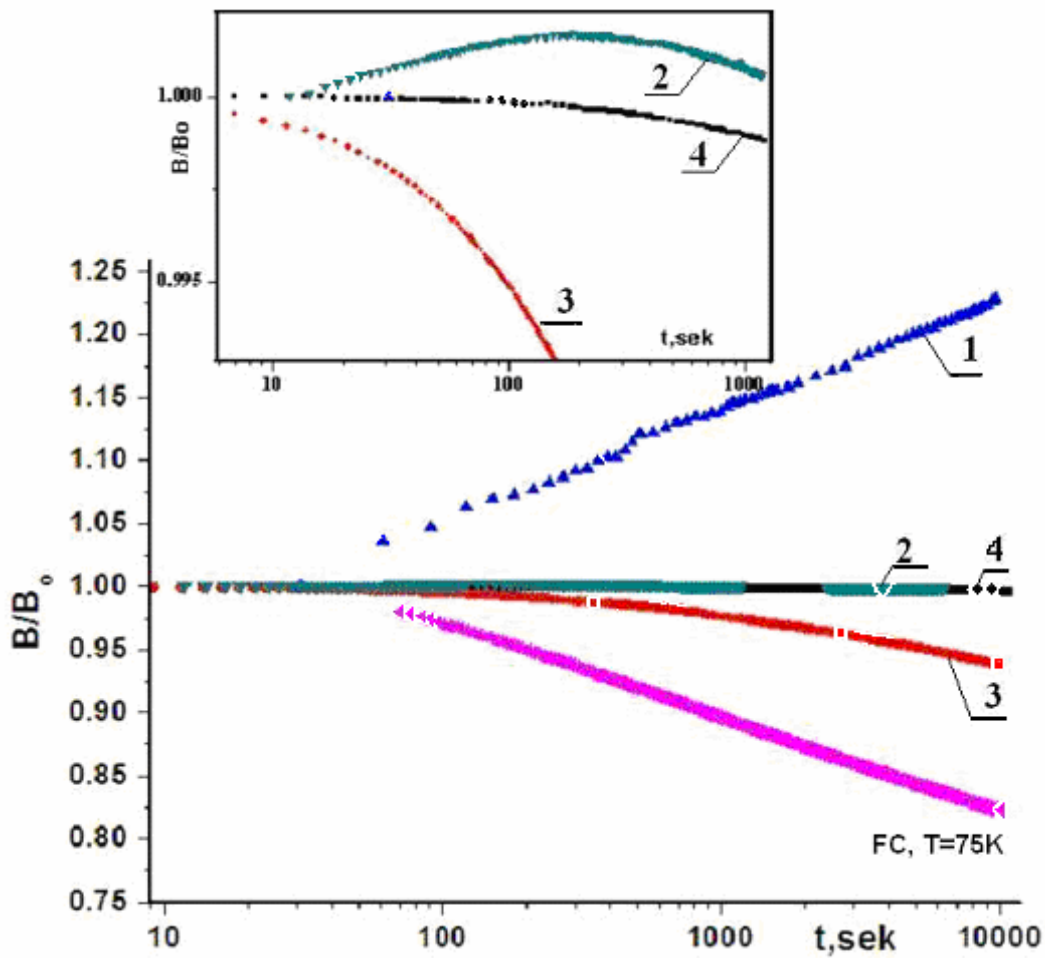

Fig. 3. Changes of magnetic flux densities corresponding to different magnitudes of the magnetization fields in time: 1(Δ)- 2.4 T, 2(∇) - 2.7T, 3(■) - 3.2T , 4(●)- 3.7 T. Symbols ◁ (lower curve) – relaxation after FC at 75 K.

Time was plotted on a logarithmic scale. It can seen that dependencies B(t) differ significantly for different states. The lowest curve in fig. 3 represents the trapped magnetic field relaxation after FC in case of an isothermal magnetization at 75 K. This curve is typical for the thermally-activated flux creep [18] and the relaxation rate is $S = -d\ln B/d\ln t = 0.032$. The upper curve (1) in fig. 3 ($\mu H_a = 2.4$T) represents an "under-magnetized" state, when the magnetic flux density at the disk center ($B_0$) is lower than at the edge. In this case B(t) increases with time. From fig. 3 one can see that the dependence B(t) is described by a logarithmic law (1), and the relaxation rate in definition (1) is negative and equal to $S = -0.037$.

Fig. 3 also contains an inset with zoomed-in initial parts of the relaxation curves. One can see that the dependence of B(t) is not monotonous near the maximum of the trapped field (curve 2 in the frame): it increases for the first 300 sec and then monotonously decreases. It is interesting to see that the relative changes $\Delta B(t)/B(0)$ are very small and over long time ($t > 10^3$ sec) the B(t) function is linear in logarithmic time scale.

Fig. 4 shows the relaxation rate (S) vs. the magnetizing field at 78 K. For non-monotonous dependences B(t) the value S was determined based on the time interval $10^3 - 10^4$ sec, where a logarithmic approximation worked well. In the "under-magnetized" state, when the value $B_0$ in the center was lower than the maximum value, the trapped flux density was increasing in time and, therefore, S was negative. When the magnitude of the magnetizing field $\mu H_a$ and, correspondingly, $B_0$ increased, the relaxation rate S decreased in absolute value and at some point changed sign. When the trapped field was maximal, (label 3 on fig. 2 and curve 3 on fig. 3) the relaxation rate was positive (S>0) and maximal as well. When the magnetization field $\mu H_a$ exceeded 3 T, the relaxation rate S decreased (along with a decrease in $B_0$) and for $\mu H_a > 3.8$ T S became a negligible number ($S < 10^{-3}$).

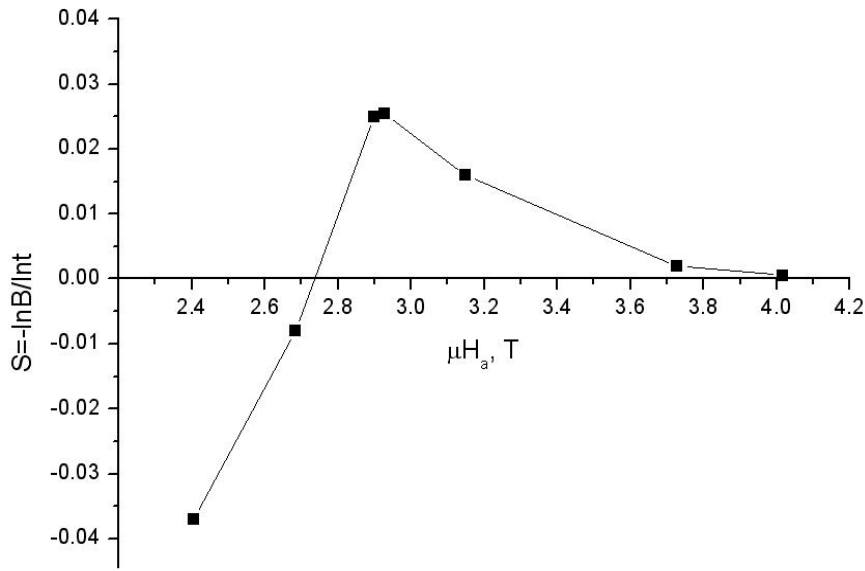

Fig. 4. Magnetic relaxation rate S as a function of the magnetizing field magnitude.

Let us summarize the main observed features of the magnetic relaxation:
- B(t) dependence has a non-monotonous character in region where S changes its sign (fig.3),
- the largest positive value of S is less than the same in "under magnetized" state and in the state FC at the same temperature.

**Discussion.**

It is well known, that in case of an isothermal magnetization a sign of S depends on the ratio between the external and internal magnetic fieds. If a superconductor was magnetized after ZFC then the external field would be higher than the internal field and dB/dr would be positive. In this case a decay of superconducting currents would lead to an increase of the internal magnetic field (S<0). If the external field was turned off after FC, then dB/dr would be negative and the trapped field would decrease in time (S>0). If a superconductor was "under magnetized" after ZFC and external field was turned off, then dB/dr would be positive at the center and negative at the periphery. In this case magnetic field at the center of superconductor would also decrease in time (H=0, S>0) because a decrease in the current density around the edge of a superconductor would reduces $B_0$ more than an increase in $B_0$ due to the relaxation at the center of superconductor where dB/dr>0. Hence at isothermal magnetization S<0 if external field H=0.

In case of a pulse magnetization, the magnetic relaxation behavior depends on the field magnitude but not the external field state (H is always zero after a pulse). A non-monotonous dependence $S(H_a)$ observed in fig.4 is explained by a non-homogeneous heating of a superconductor during the magnetic pulse when a temperature at the center of a superconductor becomes lower than a temperature at the boundary. This temperature non-uniformity results in shielding superconducting currents. Unfortunately we have no reliable data about the temperature distribution during a magnetic pulse. Instead we use estimations based on the measurement of ΔT at the center of superconductor (fig. 2). Based on the so obtained data we estimated that a change in the external magnetic field in order of 1 T would lead to a temperature jump ΔT≈1K. A non-uniform temperature distribution ΔT(r) causes a non-uniform distribution of a superconducting current density given by (3). When magnetizing is over and the sample is cooled down again to the liquid nitrogen temperature, superconducting currents will be lower than the ones that could be expected based on the critical current density at 78K. It is also known that the magnetic relaxation rate after magnetization decreases exponentially if temperature is redused [4]. For example, if the temperature was decreased by ΔT=2K from the 78K level, the creep rate would be reduced 6 times; if ΔT was 4K, the creep rate would be reduced by 200 times, and so on [4]. Because of this, currents trapped in hotter areas of a superconductor after cooling will have lower rate of relaxation than ones trapped in cooler areas. Therefore, if the local temperature initially had a radial distribution ΔT(r), then S(r) will follow the same distribution but with an exponential factor.

Analyzing position 1 on fig. 2, we know that the temperature increase at the center of the sample was less than 0.5K, and we can estimate that on edge it is was more than 2K because we know that a 2T field had penetrates into the superconductor. Therefore, we can conclude that in the central part of the sample, where dB/dr>0, the relaxation rate was significantly higher than on the boundary, and that the flux density at the center would increase in time. When $H_a$ increases, a size of the positive derivative (dB/dr>0) area decreases (the value of $r_m$ in fig. 1 decreases) and the temperature in this area increases. In this case an increase of B(t) in the central area compensates for a decrease of the flux densities in the peripheral areas. This results in a non-monotonous B(t) and a small relaxation rate.

When the magnitude of the magnetizing field exceeds $\mu H_a \approx 2.7$T, the flux density reduces from the axis to the edge (dB/dr<0) in the entire cross section of a superconductor. This corresponds to position 3 in fig. 1. In this case the currents decay leads to a decrease of $B_o$. Because a temperature increase due to the magnetization in the central area is small, the relaxation rate is largest. When the pulse magnitude increases, the superconductor temperature increases, and, consequently, the critical current density decreases in comparison to $J_c(78)$ (pos.4 on fig.1). Increasing difference $J_c(78) - J_c(T)$ is the reason why the relaxation rate decreases as the external field magnitude grows.

In a summary, in case of a pulse magnetization of superconductors a magnetic relaxation rate is shown to be a non-monotonous function of the magnetizing field strength. The density of the magnetic flux trapped in the central sample region after a pulse can either increase or decrease. In "under magnetized" state the relaxation rate can be low because of the magnetic flux compensation; however it also can be low in case of large magnetizing fields because of the inductive heating and subsequent cooling of a superconductor.

The authors are grateful to V.I.Nizhankovski (International Lab. HMF and LT in Wroclaw, Poland) for help in measuring the magnetic relaxation and A.S. Sliskov (MEPhI) for help in measuring temperature after a pulse magnetization. This work was supported by grant RFBR 06-08-01422.